\documentclass[]{spie}  
\usepackage[]{graphicx}
\def\lesssim{\mathrel{\hbox{\rlap{\hbox{\lower4pt\hbox{$\sim$}}}\hbox{$<$}}}}
\def\arcsec{\hbox{$^{\prime\prime}$}}
\def\farcs{\hbox{$.\!\!^{\prime\prime}$}}

\title{Supernovae and extragalactic astronomy with laser guide star adaptive optics} 

\author{Stuart D. Ryder\supit{a}, Seppo Mattila\supit{b},
Erkki Kankare\supit{c}, Petri V\"{a}is\"{a}nen\supit{d}
\skiplinehalf
\supit{a}Australian Astronomical Observatory, P.O. Box 915, North Ryde,
NSW 1670, Australia; \\
\supit{b}Finnish Centre for Astronomy with ESO (FINCA), University of Turku,
V\"{a}is\"{a}l\"{a}ntie 20, FI-21500 Piikki\"{o}, Finland; \\
\supit{c}Tuorla Observatory, Department of Physics and Astronomy, University
of Turku, V\"{a}is\"{a}l\"{a}ntie 20, FI-21500 Piikki\"{o}, Finland; \\
\supit{d}South African Astronomical Observatory/Southern African Large Telescope,
P.O. Box 9, 7935 Observatory, Cape Town, South Africa
}

\authorinfo{Further author information:\\
S.D.R.: E-mail: Stuart.Ryder@aao.gov.au, Telephone: $+61$ 2 9372 4843}

\begin{document} 
  \maketitle 

\begin{abstract}
Using the latest generation of adaptive optics imaging systems
together with laser guide stars on 8m-class telescopes, we are finally
revealing the previously-hidden population of supernovae in starburst
galaxies. Finding these supernovae and measuring the amount of
absorption due to dust is crucial to being able to accurately trace
the star formation history of our Universe. Our images are amongst the
sharpest ever obtained from the ground, and reveal much about how and
why these galaxies are forming massive stars (that become supernovae)
at such a prodigious rate.
\end{abstract}



\section{INTRODUCTION}
\label{s:intro}

\begin{figure}
   \begin{center}
   \begin{tabular}{c}
   \includegraphics[width=10cm]{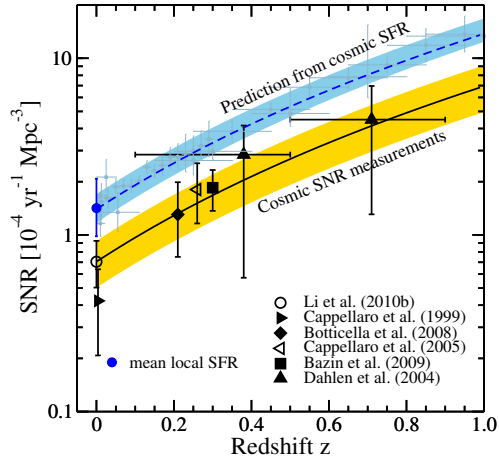}
   \end{tabular}
   \end{center}
\caption{A demonstration of the ``supernova rate problem''\cite{Horiuchi11}.
The CCSN rate
predicted from the cosmic SFR history\cite{Hopkins06} is shown by the 
light grey points and blue band, while individual determinations of the
CCSN rate at various redshifts are shown as black symbols and the yellow
band. The predicted and measured cosmic CCSN rates are consistently discrepant
by a factor of $\sim$2 beyond $z\sim0.2$.\label{f:hor}}
\end{figure}

Up to half of all the supernovae that should result when stars more
massive than $\sim$8 times the mass of our Sun reach the end of their
lives and explode go unseen. This is starkly demonstrated in
Fig.~\ref{f:hor} which compares the discovery rate of core-collapse
supernovae (CCSN) out to a redshift of $\sim$1, with the expectation
from the increasingly well-defined measures of the star formation rate
(SFR) in galaxies\cite{Horiuchi11}. This deficit of supernovae is not
for the lack of trying; many amateur and robotic supernova surveys
monitor the sky every night and, even allowing for their
incompleteness, there is a shortfall of a factor of 2 beyond our local
Universe.

Our collaboration has pioneered the use of Laser Guide Star Adaptive
Optics (LGSAO) facilities on the European Southern Observatory's Very
Large Telescope (VLT) and on the Gemini North 8 metre telescope to
begin revealing the mostly undiscovered population of CCSN within
Luminous Infra-Red Galaxies (LIRGs). Being able to detect supernovae
at infrared wavelengths in the dusty environments of LIRGs is critical
in enabling us to determine the fraction of supernovae that will be
missed by optical all-sky supernova searches such as those with the
SkyMapper, KMTNet, and Large Synoptic Survey Telescopes (LSST), as
well as satellite missions including Gaia, Euclid, and the James Webb
Space Telescope (JWST). Here we describe the objectives, the
practicalities, and the future prospects for the use of LGSAO in
finding supernovae in LIRGs, while also revealing much about the
nature of the LIRGs themselves.

\section{BACKGROUND}

\subsection{The importance of supernovae}
\label{s:sne}

Core-collapse supernovae are one of the most important phenomena in
all of astrophysics.  If massive stars did not explode having
exhausted their readily-available supplies of fuel for nuclear fusion,
then none of the newly-synthesised heavier elements locked inside
their cores would become available to the cosmos for subsequent
generations of star (and planet) formation. CCSN are among the most
energetic events in the Universe, releasing $>10^{51}$~ergs in just a
few seconds, generating shock waves which travel at tens of thousands
of kilometers per second. They give rise to exotic remnants like black
holes and neutron stars, trigger the collapse of gas clouds to form
new stars, and appear to be the underlying source of gamma ray bursts
and cosmic rays. It is crucial therefore that we have a complete
census of where and how often they occur in the Universe.

Since the most massive stars which give rise to CCSN are also the
shortest-lived, the CCSN rate is one of the most direct measures we
have of the star formation rate. Since most of the SFR diagnostics
used (e.g. H$\alpha$ and UV fluxes) and CCSN events both directly
trace massive stars, this apparent supernova deficit in
Fig.~\ref{f:hor} is not caused by uncertainty about the form or
universality of the Initial Mass Function (i.e. the relative numbers
of high- and low-mass stars formed in any burst of star
formation). Rather, the missed supernovae are either intrinsically
much dimmer than the norm, and/or hidden from our view by significant
dust extinction.

\subsection{The importance of Luminous Infra-Red Galaxies}
\label{s:lirgs}

The class of so-called Luminous Infra-Red Galaxies (LIRGs) whose
output in the infrared spectrum between 1 and 1000~$\mu$m exceeds
$10^{11}$ times that of the Sun are thought to be some of most
concentrated sites of both dust and star formation anywhere. The
absorption of ultraviolet photons from the hot young stars by the
dust, and its re-radiation at longer wavelengths, gives rise to their
copious infrared luminosity. Together with their rarer but even
brighter (by a factor of 10 or more) cousins, the Ultra-Luminous
Infra-Red Galaxies (ULIRGs), LIRGs come to dominate the total SFR over
that from normal spiral galaxies by a redshift of $\sim$1, when the
Universe was less than half its present age\cite{Magnelli09}.

On the basis of their infrared luminosity and inferred SFR, LIRGs are
expected to host in their nuclear regions about one CCSN per year on
average. Thus LIRGs ought to be ideal ``hunting grounds'' ripe for
CCSN discovery. Until quite recently only a few CCSN in LIRGs 
had been found\cite{Miluzio13}. The reasons for this are twofold:

\begin{enumerate}
\item LIRGs are typically clumpy, complex structures by contrast with
  the much smoother elliptical and spiral galaxies. Combined with the
  fact that few LIRGs are closer than 50 million parsecs away, this
  makes spotting a transient point source against such a background
  particularly challenging.
\item The same dust which so efficiently absorbs light from the young
  stars may well completely obscure the supernova's light, even when
  it briefly outshines all the other stars in the LIRG combined at
  optical wavelengths.
\end{enumerate}

\section{HUNTING FOR SUPERNOVAE IN LIRGS}

\subsection{The need for LGSAO}

The first of these factors makes spatial resolution a high priority.
Even in the best ground-based optical seeing $\sim0\farcs5$, a CCSN
within a few hundred parsecs of one of the bright nuclei of Arp~299
would not be resolvable. Fortunately Adaptive Optics (AO) facilities
such as
NaCo\footnote{http://www.eso.org/sci/facilities/paranal/instruments/naco.html}
on the VLT and
ALTAIR\footnote{http://www.gemini.edu/sciops/instruments/altair} on
the Gemini North telescope routinely deliver image quality as good as
$\sim0\farcs1$, albeit only within a small ($<30\arcsec$) field of
view, and only at near-infrared (1--2.5~$\mu$m) wavelengths.

The second factor motivates observing at near-infrared (NIR)
wavelengths rather than optical, as the extinction (usually expressed
on the logarithmic magnitude scale) due to dust is reduced by a factor
of 10.  Fortunately this is also the regime in which adaptive optics
currently performs best.

The power of NIR AO searches for CCSN was first demonstrated by us
with the discovery of SN~2004ip just 500~pc from the nucleus of the
LIRG IRAS~18293-3413\cite{Mattila07}.  The inferred extinction towards
this event was at least 5~mag in $V$, and perhaps as high as 40~mag,
meaning it would have gone unseen by any optical search. However very
few LIRGs have a nucleus which is compact enough and bright enough at
optical wavelengths to serve as the reference for on-axis AO
correction, and not many meet the alternative requirement for an
off-axis guide star. For example the ALTAIR facility on Gemini North
requires that a $V<11$ guide star be located within $25\arcsec$ of the
target for full AO correction, or $V<15$ for partial correction. Many
of the most luminous LIRGs, which potentially offer the highest CCSN
rates, simply have no suitable natural guide stars
available. Fortunately the advent of LGSAO facilities on 8m class
telescopes a decade ago relaxed this requirement somewhat, opening up
many more LIRGs for CCSN searches. Nevertheless ALTAIR in LGSAO mode
still requires a $R<18$ tip/tilt star within $25\arcsec$ of the target
for a low Strehl (up to 10\% in $K$) correction.

\subsection{Observing strategy}

As mentioned previously, spotting a new point source like a CCSN
against a complex background such as a LIRG is quite challenging.  The
standard approach to CCSN discovery is to subtract off a prior
``reference'' image of the LIRG host from the latest image (preferably
with both images obtained by the same facility). Before subtraction
the latest image must be rotated, shifted, and possibly scaled to
match the earlier image using isolated field stars or other compact
sources as a reference.  The image with the better image quality as
defined by the point spread function (PSF) width must be smoothed with
a convolution kernel to match the image quality achieved in the other
image, and the sky backgrounds and flux scales also adjusted to
match. We use a slightly modified version of the Optimal Image
Subtraction method\cite{Alard00} as implemented in ISIS
2.2\footnote{http://www2.iap.fr/users/alard/package.html}.

Any new point source residual in the pair-subtracted image is merely
a CCSN candidate as other possibilities include:
\begin{itemize}
\item a passing minor planet. At AO resolutions the proper motion of a
  minor planet is usually detectable between, or even within
  successive exposures.
\item variable foreground stars, or variability of an Active Galactic
  Nucleus (AGN). The former are unlikely due to the typically small
  field of view, and either would show light curves compiled from
  follow-up observations that are inconsistent with that expected for
  a CCSN.  Furthermore LIRGs can be selected on the basis of their
  {\em IRAS} colors to be dominated by star formation, rather than by
  an AGN.
\item a thermonuclear (Type Ia) supernova. These events are thought to
  primarily trace lower mass star formation, with a much longer time
  delay ($\sim10^{9}$~years). We expect perhaps 5\% of all supernovae
  detected in LIRGs to be Type Ia and not a CCSN\cite{Mattila07}, and
  a radio detection would rule out a Type Ia event\cite{Hancock11}.
\end{itemize}

By convention the International Astronomical Union's Central Bureau
for Astronomical Telegrams (CBAT) requires that before conferring an
official designation, any CCSN candidate must be independently
confirmed both by imaging on a different night and/or at a different
facility, as well as by optical spectroscopy that allows assignment of
a sub-type (e.g. Type Ia, Ib, Ic, IIP, IIL, IIb, IIn, etc.). Both of
these requirements are problematic in our case, since LGSAO facilities
are few in number and hard to win time on, while some of the CCSN
candidates are so heavily extincted by dust that obtaining an optical
spectrum is impossible. Instead we make use of the fact that infrared
emission from a CCSN tends to follow distinctive patterns\cite{Mattila01}.
Furthermore, by
monitoring how the flux of the CCSN evolves over many months in the 3
NIR bands $J$ (centred on 1.25~$\mu$m), $H$ (1.65~$\mu$m), and $K$
(2.20~$\mu$m) and fitting these to templates of CCSN suffering almost
no extinction, it is possible to derive the line-of-sight extinction
due to dust, which is one of the key goals of our program. By
measuring the extinction distribution in CCSN which are detected at
NIR wavelengths, we can more realistically model the fraction of all
CCSN which will be missed at optical wavelengths.

The last key to a successful CCSN survey is the cadence of
observation, i.e. for a given amount of total observing time, what is
the optimal interval between repeat observations of a LIRG? Observing
each LIRG at monthly intervals say would almost guarantee that no CCSN
bright enough at peak to be detected would be missed, but would
rapidly exhaust all the available observing time. On the other hand,
observing each LIRG only once per year would allow ample time for a
CCSN event to peak and then fade from view in between observing
epochs. Our simulations\cite{Mattila01,Mattila04} indicate that
observations spaced 3--4 months apart yield the most efficient use of
observing time, while minimising the risk of missing a CCSN event.
Such a cadence favors observations in a queue mode, as the cost of so
many visits (approximately monthly to monitor a sufficiently large
number of LIRGs) to an LGSAO facility would be prohibitive.

\subsection{Supernova discoveries}

Over the course of 8~semesters between 2008 and 2012 we were allocated
just over 50~hours with the ALTAIR LGSAO facility and Near InfraRed
Imager (NIRI) on the Gemini North 8~m telescope on Mauna Kea to
monitor 8~LIRGs for new CCSN. Each LIRG was imaged on average 7 times,
at intervals of between 1 and 12 months but on average every
4.5~months. Each epoch of observation consisted of $9\times$30~sec
dithered exposures on-source, followed by an equivalent set of
exposures on adjacent blank sky as the LIRGs filled most of the NIRI
field of view. The total elapsed time for each epoch including
overheads was typically less than half an hour.

\begin{figure}
   \begin{center}
   \begin{tabular}{c}
   \includegraphics[width=15cm]{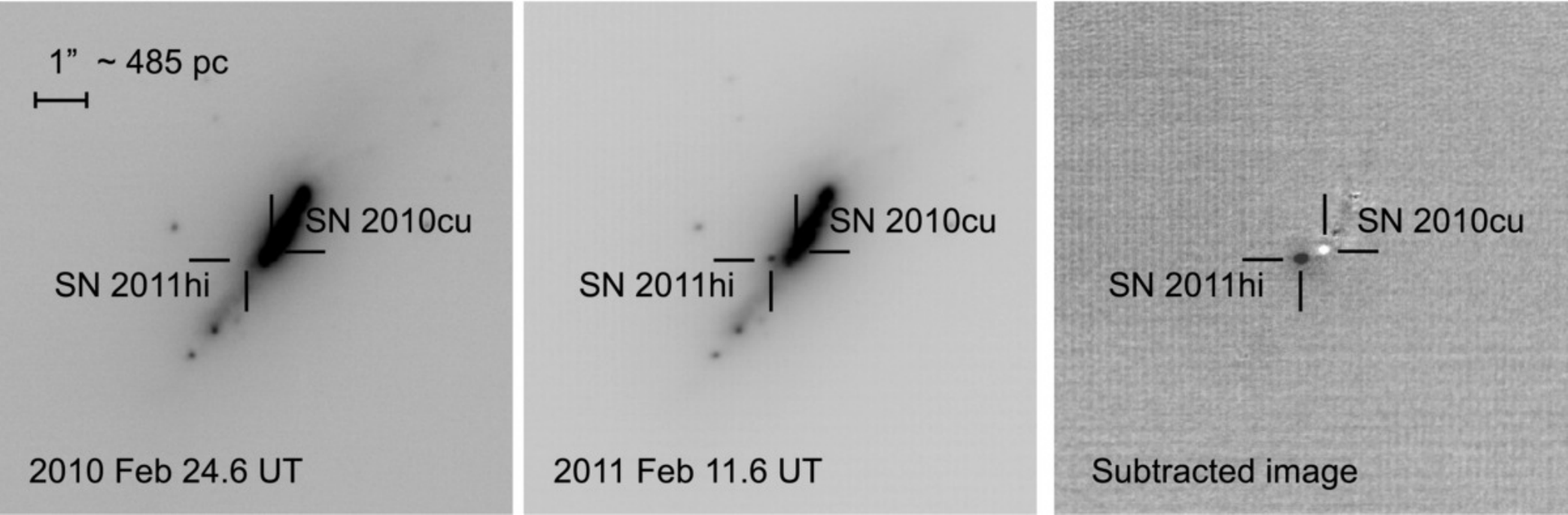}
   \end{tabular}
   \end{center}
\caption{$10\arcsec \times 10\arcsec$ subsections of $K$-band Gemini
  ALTAIR/NIRI LGSAO discovery images of SN~2010cu (left panel) and
  SN~2011hi (middle panel) and the subtraction between these two
  (right panel)\cite{Kankare12}. The smooth subtraction of the host
  galaxy IC 883 clearly demonstrates the good alignment and PSF match
  between the images and the two SNe can be clearly detected as
  individual point sources thanks to the high-angular resolution
  provided by LGSAO. North is up and east is to the
  left.\label{f:ic883}}
\end{figure}

At the conclusion of our survey we discovered 4 new CCSN, and
confirmed 2 more (Table~\ref{t:sne}) which had been detected
independently. Just as importantly, we succeeded in obtaining $J$,
$H$, and $K$ light curves and/or radio
follow-up\cite{Romero12,Romero14} for nearly all of them, to ascertain
their line-of-sight extinctions, and confirm their core-collapse
nature, respectively. The range in extinction probed ranges from
effectively none, up to the equivalent of almost 20~magnitudes in the
$V$ band (i.e. only 1 in 40~million optical photons would make it
through the dust). Figure~\ref{f:ic883} illustrates how one discovery
can flow from the next, in that follow-up imaging of our discovery
SN~2010cu in IC 883 yielded the discovery of SN~2011hi, the second
CCSN in IC~883 within a year.

\begin{table}[h]
\caption{Summary of supernovae discovered, or confirmed (SN 2010O, SN 2010P)
by our Gemini North/ALTAIR laser guide star program, including line-of-sight
extinction and projected distance from the LIRG nucleus.\label{t:sne}}
\begin{center}
\begin{tabular}{llccc}
\hline
Supernova & LIRG Host      & Extinction   & Projected      & Reference \\
          &                & A$_{V}$ (mag) & distance (pc)  &       \\
\hline
SN 2004iq & IRAS 17138-1017    & 0--4    &  700 &  \citenum{Kankare08} \\
SN 2008cs & IRAS 17138-1017    & 17--19  & 1500 &  \citenum{Kankare08} \\
SN 2010O  & IC 694 (Arp 299)   & 2       & 1100 &  \citenum{Kankare14} \\
SN 2010P  & NGC 3690 (Arp 299) & 7       & 1200 &  \citenum{Kankare14} \\
SN 2010cu & IC 883             & 0--1    &  200 &  \citenum{Kankare12} \\
SN 2011hi & IC 883             & 5--7    &  360 &  \citenum{Kankare12} \\
\hline
\end{tabular}
\end{center}
\end{table}

\begin{figure}
   \begin{center}
   \begin{tabular}{c}
   \includegraphics[width=12cm]{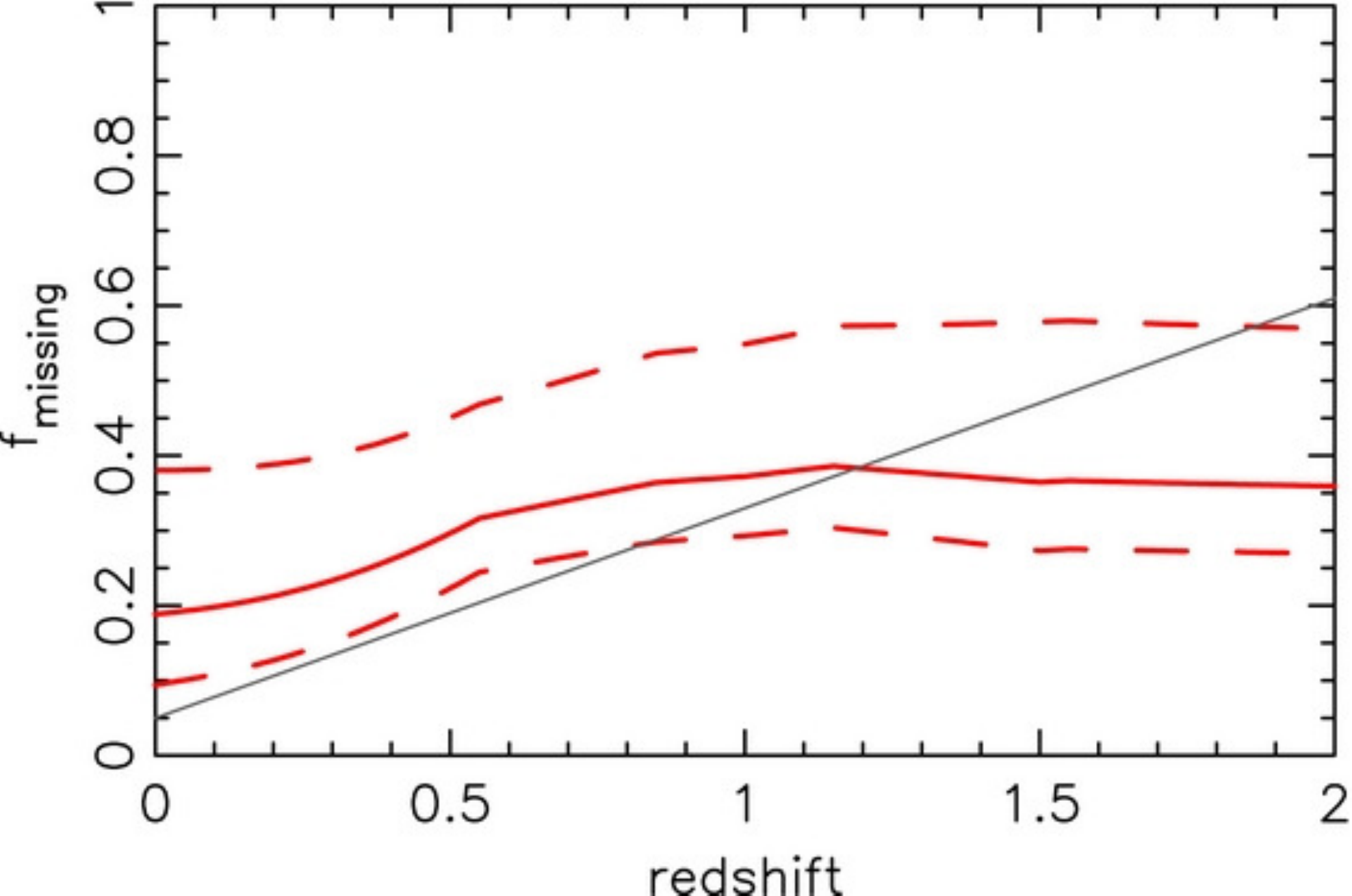}
   \end{tabular}
   \end{center}
\caption{The fraction of CCSN missed by rest-frame optical searches
  as a function of redshift\cite{Mattila12}. Red lines show our best
  estimate together with the uncertainties indicated by the dashed
  lines. The solid grey line is the previous best estimate for the
  missing fraction\cite{Mannucci07}.\label{f:sm12}}
\end{figure}

Our intensive monitoring of the LIRG Arp~299 has enabled us to
build upon earlier efforts\cite{Mannucci07} to derive the fraction of
CCSN missed by optical surveys as a function of
redshift\cite{Mattila12} (Fig.~\ref{f:sm12}), which as mentioned in
Section~\ref{s:intro} has implications for current and future CCSN
surveys, and even for the diffuse supernova neutrino
background\cite{Lien10}. Our corrections for the missing fractions of
CCSN have been employed\cite{Dahlen12,Melinder12} to show that most,
if not all of the missing supernovae might be accounted for by dust
extinction alone. But at low redshift statistical errors still
dominate, so we need to increase the number of known CCSN in LIRGs
before we can begin to discriminate between dusty and intrinsically
dim CCSN.

\section{THE NATURE OF LIRGS}

In addition to the CCSN discoveries we have made, one of the
significant legacies that our work has delivered are some of the
sharpest, deepest images of LIRGs ever obtained from the ground.
While the triggering mechanism for prodigious star formation in LIRGs
is believed to be mergers and interactions between two or more
galaxies concentrating significant quantities of gas into their
nuclei, very few of the LIRGs in our CCSN sample have been studied in
any great detail individually. We have been using the best images from
our Gemini/ALTAIR sample, as well as from an earlier VLT/NaCo survey,
to glean new insights into the nature of the LIRGs themselves.

\subsection{A triple merger}

\begin{figure}
   \begin{center}
   \begin{tabular}{c}
   \includegraphics[width=12cm]{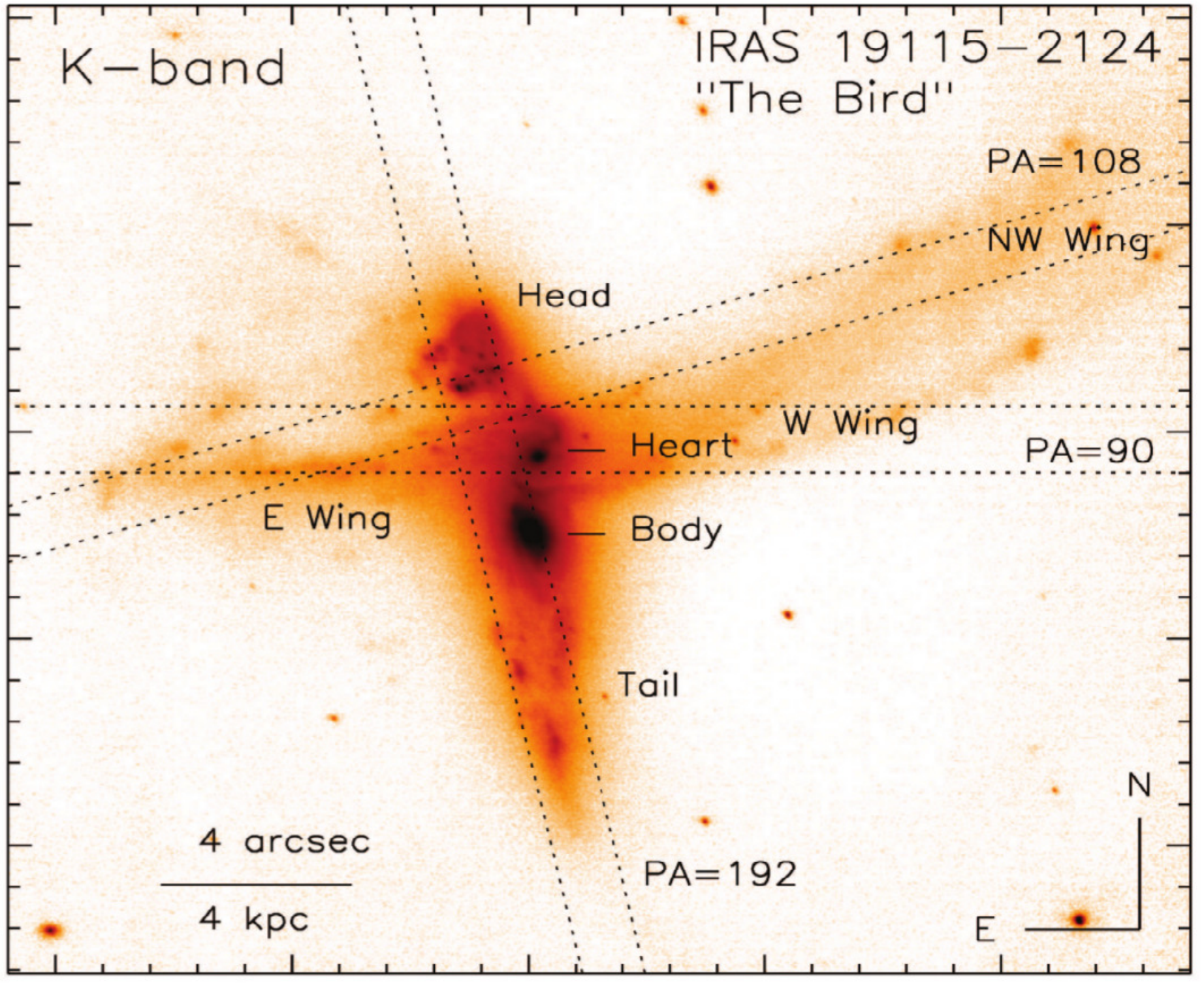}
   \end{tabular}
   \end{center}
\caption{VLT/NaCo natural guide star AO image of the ``Bird'', with the
primary components indicated.\label{f:bird}}
\end{figure}

The LIRG IRAS 19115−2124 has been dubbed the ``Bird'' on account of
its shape, which in NaCo $K$-band images resolves into two wings, a
head, body, heart, and extended tail\cite{Vaisanen08a}
(Fig.~\ref{f:bird}).  By combining this image with data from the {\em
  Hubble Space Telescope} (HST) and {\em Spitzer} satellite, as well
as optical longslit spectroscopy with the Southern African Large
Telescope, we reach the surprising conclusion that the Bird contains
not two, but three galaxies undergoing mutual interaction. The head,
heart, and body are each separate galaxies of between 1 and
$7\times10^{10}$ solar masses, with the wings marking tidal
tails. Strangely, despite being the least massive of the three
galaxies, it is the head which is currently dominating star formation
in the Bird.

\subsection{Leading spiral arms}

\begin{figure}
   \begin{center}
   \begin{tabular}{c}
   \includegraphics[width=15cm]{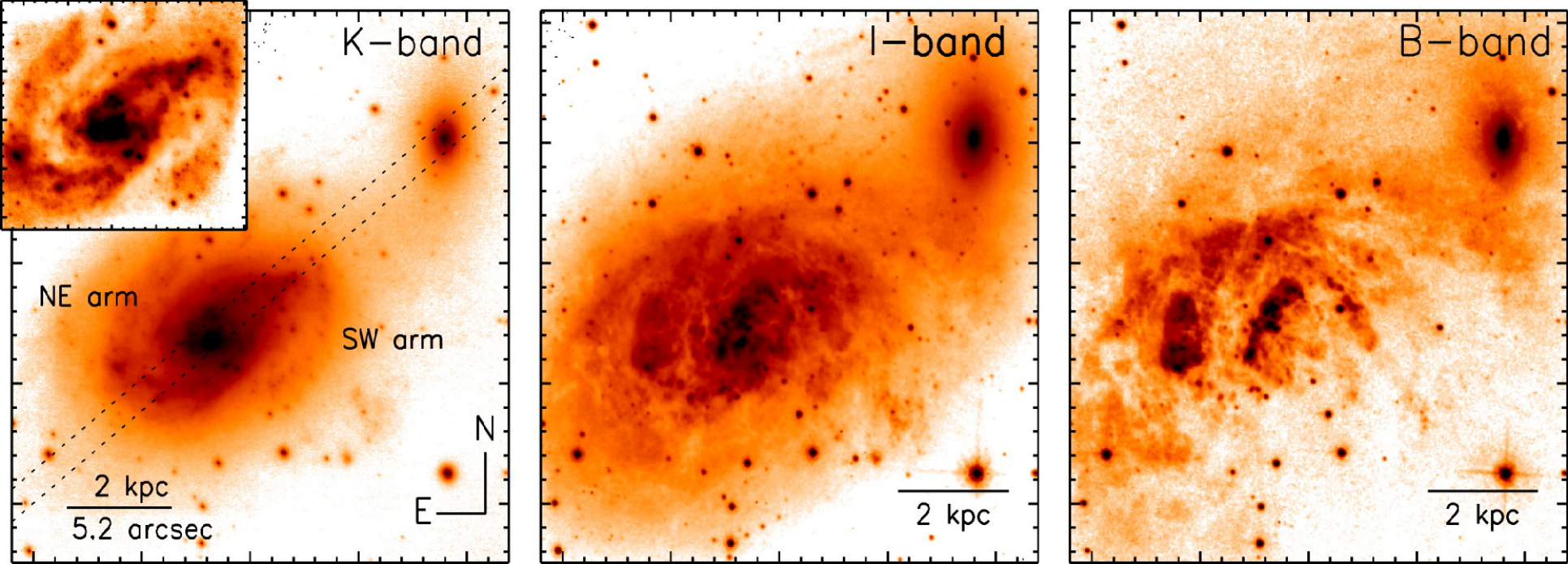}
   \end{tabular}
   \end{center}
\caption{VLT/NaCo natural guide star AO image of the LIRG IRAS 18293-3413
({\em left}), with the inset showing an unsharp masked image to highlight
the arms in better contrast. The HST $I$-band ({\em middle}) and $B$-band
({\em right}) images are also shown.\label{f:i18293}}
\end{figure}

The LIRG IRAS 18293-3413 not only hosted our first AO CCSN discovery
SN~2004ip\cite{Mattila07}, but also turns out to have unusual
dynamics. HST optical images in Figure~\ref{f:i18293} show
significantly more dust in silhouette towards the southwest,
consistent with this edge of the galaxy disk being closer to us than
the northeast edge. NIR longslit spectroscopy with the IRIS2
instrument on the Anglo-Australian Telescope indicates that the
northwest side of the disk, facing the companion galaxy, is
approaching us. Putting these two facts together we are led to
conclude\cite{Vaisanen08b} that the galaxy is turning clockwise in the
image, i.e., in the same direction that the spiral arms clearly
visible in the inset to Fig.~\ref{f:i18293} open out. Thus IRAS
18293-3413 is one of just a handful of known or suspected ``leading
arm'' spirals. Such leading arms are predicted in simulations of
retrograde encounters of a small companion galaxy\cite{Byrd93} as seen
in IRAS~18283-3413.

\subsection{Super star clusters}
\label{s:sscs}

AO images of LIRGs show them to be extremely rich in what are referred
to as ``super star clusters'' (SSCs; Fig.~\ref{f:sn13xx}) having
masses between $10^{5}$ and $10^{7}$ solar masses, ages of 10--100
million years, and sizes of just 3--5~parsecs. Analysis of the
$K$-band luminosity functions of the SSC populations in our
LIRGs\cite{Randriamanakoto13a} shows a power law with values of the
index $\alpha$ ranging between 1.5 and 2.4 with an average value of
1.9, which is less steep than the average of 2.2 in normal spiral
galaxies. Furthermore we have been able to
demonstrate\cite{Randriamanakoto13b} that the luminosity of the
brightest SSC in a LIRG scales with the galaxy's total star formation
rate, but with a steeper slope than a single optical to NIR conversion
would imply.

\section{THE FUTURE}

To provide more meaningful constraints on the missing CCSN fraction we
need to discover many more CCSN and have expanded our efforts to the
Southern hemisphere making use of the superior new Gemini
Multi-conjugate AO System (GeMS) feeding the Gemini South AO Imager
(GSAOI) with its $85\arcsec \times 85\arcsec$ field of view. We
anticipate that the ability of GeMS to deliver a uniform PSF across
the full GSAOI field will make for much improved image subtraction and
photometric calibration compared with the ``classical'' LGSAO systems
used to date, and potentially allow us to probe for CCSN even closer
to the bright LIRG nuclei.

\begin{figure}
   \begin{center}
   \begin{tabular}{c}
   \includegraphics[width=16cm]{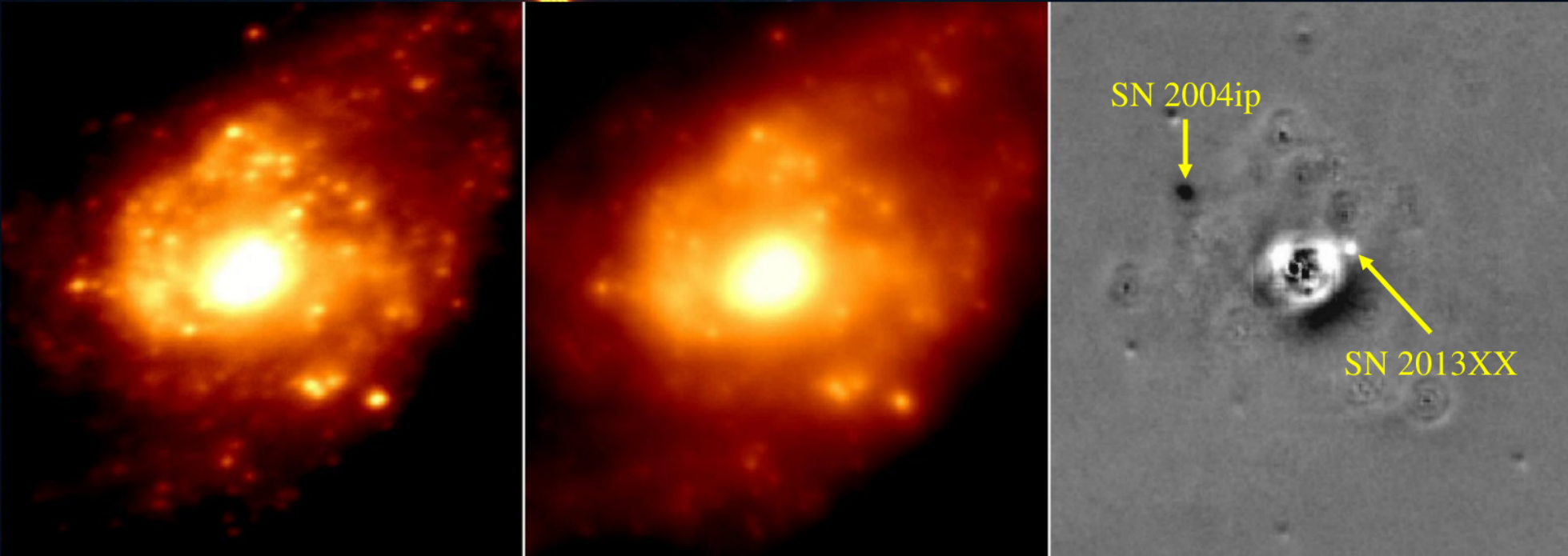}
   \end{tabular}
   \end{center}
\caption{({\em left}) $K_{\rm s}$-band GeMS/GSAOI image of the LIRG
  IRAS~18293-3413 from 21~April 2013; ({\em middle}) a VLT NaCo image
  of the same LIRG from 2004; ({\em right}) an optimal image
  subtraction of the NaCo image from the GeMS image. The negative
  residual to the northeast is SN 2004ip discovered by us in our
  earlier NaCo program\cite{Mattila07} but which has long since faded
  from view; the positive residual $\sim0\farcs5$ ($\sim$200~parsecs
  projected distance) to the northwest is a new supernova revealed in
  the GeMS image marked here as ``SN2013XX''.
  Notice also the
  increase in the number of SSCs (Sec.~\ref{s:sscs}) apparent in the
  new GeMS/GSAOI image due to the improvement in image
  quality.\label{f:sn13xx}}
\end{figure}

As an illustration of this potential, Fig.~\ref{f:sn13xx} shows our
first epoch image with GeMS/GSAOI of the LIRG IRAS 18293-3413,
compared with our previous best image from NaCo. In addition to
highlighting the advance in AO performance in the past decade, these
data nicely illustrate both our very first discovery of a CCSN with
AO, SN~2004ip\cite{Mattila07}, as well as what we believe to be our
first CCSN discovery with GeMS/GSAOI,
marked here with the unofficial designation ``SN~2013XX''.
Thus we are confident that the next generation of LGSAO
will leave little room for core-collapse supernovae to remain hidden
within Luminous Infrared Galaxies.

\acknowledgments 
 
We are grateful to all our collaborators in this work for their
contributions, in particular Cristina Romero-Ca\~{n}izales, Miguel
P\'{e}rez-Torres, Jari Kotilainen, and Zara Randriamanakoto. The efforts
of Gemini and VLT staff in developing and successfully operating
common-user, queue-scheduled LGSAO facilities in the face of all kinds
of challenges (not the least of which in Gemini's case is having to
obtain laser propagation clearance from the US Space Command) are
deeply appreciated. This work is based on observations obtained at the
Gemini Observatory, which is operated by the Association of
Universities for Research in Astronomy, Inc., under a cooperative
agreement with the NSF on behalf of the Gemini partnership: the
National Science Foundation (United States), the National Research
Council (Canada), CONICYT (Chile), the Australian Research Council
(Australia), Minist\'{e}rio da Ci\^{e}ncia, Tecnologia e
Inova\c{c}\~{a}o (Brazil) and Ministerio de Ciencia, Tecnolog\'{i}a e
Innovaci\'{o}n Productiva (Argentina).



\newpage
\bibliography{ryder} 
\bibliographystyle{spiebib} 

\end{document}